\documentclass[pdftex, pra, twocolumn, showpacs, floatfix, superscriptaddress]{revtex4}

\usepackage{times}
\usepackage{amsfonts}
\usepackage{amssymb}
\usepackage{amsmath}
\usepackage{graphicx}
\usepackage[colorlinks=true, breaklinks=true, linkcolor=blue, citecolor=blue, urlcolor=blue]{hyperref}

\newcommand{\doi}[2]{\href{http://dx.doi.org/#1}{#2}}

\begin{document}

\title{Self-bound many-body states of quasi-one-dimensional dipolar Fermi gases: \\ Exploiting Bose-Fermi mappings for generalized contact interactions}

\author{F. Deuretzbacher}
\email{frank.deuretzbacher@itp.uni-hannover.de}
\affiliation{Institut f\"ur Theoretische Physik, Leibniz Universit\"at Hannover, Appelstr. 2, DE-30167 Hannover, Germany}

\author{G. M. Bruun}
\affiliation{Department of Physics and Astronomy, University of Aarhus, Ny Munkegade, DK-8000 Aarhus C, Denmark}

\author{C. J. Pethick}
\affiliation{The Niels Bohr International Academy, The Niels Bohr Institute, University of Copenhagen, Blegdamsvej 17, DK-2100 Copenhagen \O, Denmark}
\affiliation{NORDITA, KTH Royal Institute of Technology and Stockholm University, Roslagstullsbacken 23, SE-10691 Stockholm, Sweden}

\author{M. Jona-Lasinio}
\affiliation{Institut f\"ur Theoretische Physik, Leibniz Universit\"at Hannover, Appelstr. 2, DE-30167 Hannover, Germany}

\author{S. M. Reimann}
\affiliation{Mathematical Physics, LTH, Lund University, SE-22100 Lund, Sweden}

\author{L. Santos}
\affiliation{Institut f\"ur Theoretische Physik, Leibniz Universit\"at Hannover, Appelstr. 2, DE-30167 Hannover, Germany}

\begin{abstract}

Using a combination of results from exact mappings and from mean-field theory we explore the phase diagram of quasi-one-dimensional systems of identical fermions with attractive dipolar interactions. We demonstrate that at low density these systems provide a realization of a single-component one-dimensional Fermi gas with a generalized contact interaction. Using an exact duality between one-dimensional Fermi and Bose gases, we show that when the dipole moment is strong enough, bound many-body states exist, and we calculate the critical coupling strength for the emergence of these states. At higher densities, the Hartree--Fock approximation is accurate, and by combining the two approaches we determine the structure of the phase diagram. The many-body bound states should be accessible in future experiments with ultracold polar molecules.

\end{abstract}


\maketitle

\section{Introduction}

One of the advances in exactly solvable quantum mechanical models in recent years has been the demonstration that there is a duality between a single-component one-dimensional (1D) Fermi gas with short-range ``$p$-wave'' interactions and a Bose gas with a contact interaction~\cite{Cheon99, Granger04}. This enables one to map the fermion problem to a boson one, which has been solved exactly~\cite{Lieb63, McGuire64}. Two-component Fermi gases with a short-range tunable interaction have been realized with atomic gases and this has led to fundamental discoveries~\cite{Giorgini08}. However, the Pauli principle prohibits any significant interaction effects in a single-component gas of fermionic atoms, which therefore is essentially noninteracting. 

We argue in this article that low-density single-component quasi-1D systems of fermions with dipolar interactions provide a realization of a Fermi gas with a generalized contact interaction. Dipolar interactions, particularly those between electric dipole moments, have a number of important features: they can be strong, they can be tuned by varying the strength and direction of an aligning electric field, and they are anisotropic. As a consequence, dipolar gases exhibit qualitatively new physics which has been extensively investigated in recent years~\cite{Baranov08,Lahaye09,Baranov12}. Of particular importance are heteronuclear diatomic molecules, which have appreciable electric dipole moments, and many groups are currently working on cooling such molecules to quantum degeneracy~\cite{Ni08, Chotia12, Wu12, Heo12}.

Here we focus on quasi-1D systems of single-component fermions with attractive dipolar interactions that can be realized by aligning dipoles along the length of the system. First, we show that for low particle densities the dipolar interaction behaves as a generalized contact interaction. Therefore the properties of many-body systems can be calculated using the boson-fermion duality established in Refs.~\cite{Cheon99, Granger04}. By these methods we show that the system has many-body bound states for a sufficiently strong dipolar interaction. For higher particle densities, the nonzero range of the interaction must be taken into account and we calculate properties from mean-field theory: we thereby delineate the coupling constants for which self-bound states exist and calculate their density. The density of these self-bound states is determined by the short-range length scale of the interaction.

\section{Effective 1D interaction}

We consider a system of dipoles with mass $m$ translationally invariant in the $z$ direction and confined in the transverse directions by a harmonic potential with oscillation frequency $\omega_\perp$. For this situation, the effective 1D potential for dipoles oriented in the $z$-direction and in the lowest energy state of the transverse motion is obtained by averaging the dipolar interaction over the ground state of the transverse motion and is given by~\cite{Sinha07, Deuretzbacher10, Bartolo13}
\begin{equation} \label{Vdd}
V(z) = - \frac{d^2}{l_\perp^3} \int_0^\infty dw \, w^2 e^{-w^2 / 2 - w |z| / l_\perp} ,
\end{equation}
with $l_\perp = \sqrt{\hbar / m \omega_\perp}$ and $d^2$ the strength of the dipolar interaction, which is given by $d^2 = D^2 / (4 \pi \epsilon_0)$ in the case of electric dipoles and by $d^2 = \mu_0 g_L^2 \mu_B^2 / (4 \pi)$ in the case of magnetic dipoles. Here, $D$ is the electric dipole moment, $\epsilon_0$ is the electric constant, $\mu_0$ is the magnetic constant, $g_L$ is the Land\'e factor, and $\mu_B$ is the Bohr magneton. For $|z| \gg l_\perp$,  $V(z) \simeq -2 d^2 / |z|^3$, corresponding to the bare dipole interaction, which is attractive since the dipoles are oriented along the $z$ axis, while for $|z| \rightarrow 0$, ${V(z) \rightarrow V(0) = -\sqrt{\pi/2} \, d^2 / l_\perp^3}$, which is finite. A key point is that the $1/|z|^3$ behavior means that the interaction behaves as a short-range potential with a range $\sim l_\perp$.

\section{Two-body problem}

\begin{figure}
\begin{center}
\includegraphics[width =\columnwidth]{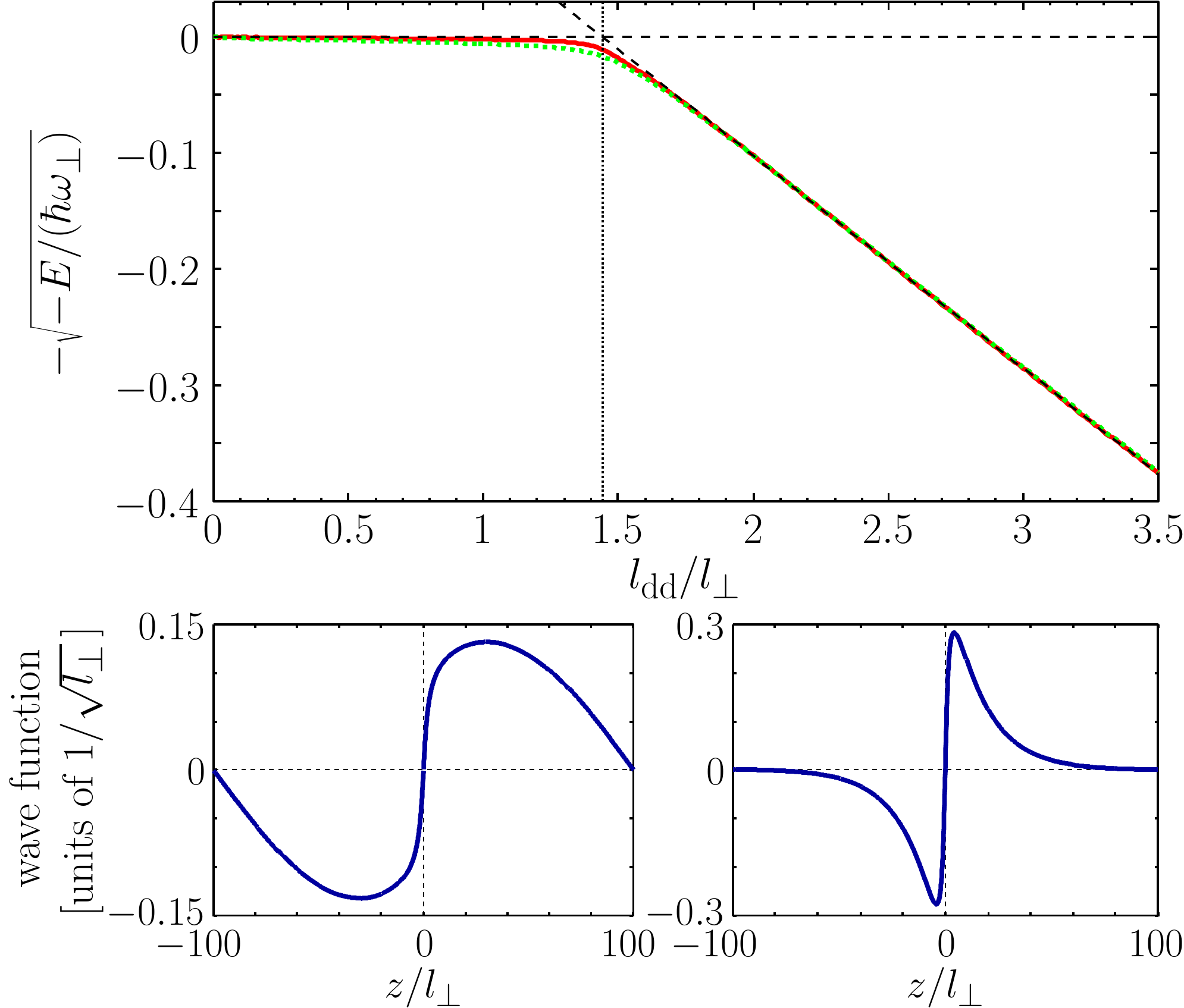}
\caption{(Color online) Top: Plot showing the dependence of the energy $E$ of the lowest two-body state on the strength of the attractive interaction, $l_\text{dd} / l_\perp$, for two box lengths: $L / l_\perp = 400$ [dotted (green) line] and $800$ [solid (red) line]. To bring out the absence of binding for $l_\text{dd} < l_\text{crit} \approx 1.44 \, l_\perp$ and the expected $(l_\text{dd} - l_\text{crit})^2$ behavior of the binding energy near threshold, we plot the quantity $-\sqrt{-E/(\hbar \omega_\perp)}$. The dashed line is a fit of the numerical results to a linear function $\propto l_\text{dd} - l_\text{crit}$, and this holds over a wider range of couplings than might naively be anticipated. The rounding of the plots for $l_\text{dd} \approx l_\text{crit}$ is due to the finite size of the box. Bottom: Relative wave function for $l_\text{dd} / l_\perp = 1.3$ (left) and $1.8$ (right) and box length 200$\, l_\perp$.}
\label{fig-2body}
\end{center}
\end{figure}

To begin with, let us consider the problem of two identical fermionic dipoles in one dimension interacting via the potential,~(\ref{Vdd}). (The two-boson problem is discussed in Ref.~\cite{Bartolo13}.) For identical fermions, the relative wave function must be odd in $z$ and we have calculated the energy of the lowest state by solving the 1D Schr\"odinger equation numerically. The strength of the dipolar coupling is conveniently measured in terms of the length $l_\text{dd} = m d^2 / \hbar^2$. Figure \ref{fig-2body} (top) shows the energy of the lowest state of the relative motion which is assumed to be confined to boxes of length $400$ [dotted (green) line] and $800 \, l_\perp$ [solid (red) line] as a function of $l_\text{dd}$. For $l_\text{dd} > l_\text{crit} \approx 1.44 \, l_\perp$ there is a bound state and the leading contribution to the binding energy varies as  $(l_\text{dd} - l_\text{crit})^2$. The relative wave function for two cases, one corresponding to a state in the continuum (bottom left) and the other to a bound state (bottom right), are shown when the relative motion is confined to a box of length $200 \, l_\perp$. Both wave functions change rapidly within the range of the interaction and comparatively slowly outside. Outside the range of the potential, but for distances not too close to $L/2$, the wave function of a bound state of the relative motion has the form
\begin{equation}
\psi(z) \propto \text{sgn}(z) e^{-|z|/\rho} ,
\end{equation}
where $z$ is the relative coordinate and $\rho = \hbar / \sqrt{m E_B}$, $E_B$ being the binding energy of the bound state.

\section{Insights from exactly soluble models}

To construct the many-body wave function at low density, we use the fact that the interaction has a short range. When the particles are outside the range of the interaction, the problem is then equivalent to that of identical fermions interacting via a ``$p$-wave'' contact interaction~\cite{Cheon99, Granger04}. This is extremely useful since we can then use exact results for 1D systems. When the interaction is sufficiently strong that a bound state for two fermions exists, the fermionic many-body system with a $p$-wave contact interaction can be mapped to a bosonic one with a usual ($s$-wave) contact interaction of strength $g_\text{1D}^B = -2 \hbar \sqrt{E_B / m}$~\cite{Cheon99, Granger04}. The exact many-body ground state of the bosonic system is known~\cite{McGuire64}, and using the mapping of Refs.~\cite{Cheon99, Granger04} we obtain the $N$-body wave function for fermions in the region outside the range of the interaction, i.e., for $|z_i-z_j| \gg l_\perp$ (for all pairs $i, j$), which is given by
\begin{equation} \label{psiout}
\psi \propto \prod_{i<j} \text{sgn}(z_i-z_j) e^{-|z_i-z_j| / \rho}.
\end{equation}
The Bose system is self-bound and at the center of mass of the particles, the 1D particle density is $n(0) \propto N^2 / \rho$~\cite{Calogero75}, which diverges for $N \rightarrow \infty$.

The dipolar system is saved from this collapse in the thermodynamic limit by the nonzero range $\sim l_\perp$ of the interaction,~(\ref{Vdd}). Indeed, the approximation of treating the interaction as being of zero range is good only if the separation between particles is large compared with the range of the interaction, $\sim l_\perp$. In particular, the wave function,~(\ref{psiout}), will be a poor approximation if $n \gtrsim 1 / l_\perp$, and the predicted collapse of the system to a density $\sim N^2 / \rho$ is an artifact due to the failure of the contact interaction assumption.

Nevertheless, it is still possible to demonstrate that there is a many-body bound state when there is a two-body bound state. For low densities with a particle spacing much larger than $l_\perp$, the mapping of the dipolar system onto a Bose gas with an attractive contact interaction is accurate. From this it immediately follows that the energy per particle is $g_\text{1D}^B n / 2$. Thus, if $g_\text{1D}^B < 0$, i.e., $l_\text{dd} \geqslant l_\text{crit}$, the energy decreases linearly with increasing density $n$ and there is a many-body bound state. This linear decrease is shown by the dashed lines in Fig.~\ref{fig-Nbody}, where we plot the energy per particle as a function of the density for various coupling strengths $l_\text{dd} \geqslant l_\text{crit}$. The low density state is, in general, not an equilibrium one, since the energy is not a minimum with respect to variations of $n$. With a contact interaction, the system would collapse to a density of order $N^2 / \rho$, but when the short-range behavior of the potential is taken into account, the equilibrium density will be set by the short-range scale of the interaction, which is independent of $N$ in the limit of a large number of particles, i.e., $n \sim 1 / l_\perp$.

If the dipolar coupling is insufficient to create a two-body bound state with odd symmetry, the coupling constant $g_\text{1D}^B$ in the analogous boson problem is positive, and the solution of the many-body problem is that obtained by Lieb and Liniger~\cite{Lieb63}. For $n \ll m g_\text{1D}^B / \hbar^2$ the system behaves like a noninteracting Fermi gas and the energy per particle is $E/N \lesssim \hbar^2 \pi^2 n^2 / (6m)$, while for $n \gtrsim m g_\text{1D}^B / \hbar^2$ the system resembles a weakly interacting Bose gas and the energy per particle is $E/N \lesssim g_\text{1D}^B n / 2$. The lowest energy state has a uniform density, and the pressure is positive.

[The strength of the $s$-wave contact interaction of the boson problem, obtained from a fit to the two-body binding energy, is given by $g_\text{1D}^B = 0.37 (1.44 - l_\text{dd}/l_\perp) \hbar \omega_\perp l_\perp$. The linear dependency of $g_\text{1D}^B$ on $(l_\text{crit} - l_\text{dd})/l_\perp$ is also valid for $l_\text{dd} \lesssim l_\text{crit}$. Therefore, the system behaves like a noninteracting Fermi gas, if $n \, l_\perp \ll m g_\text{1D}^B l_\perp / \hbar^2 = 0.37 (1.44 - l_\text{dd}/l_\perp)$, and like a weakly interacting Bose gas, if $n \, l_\perp \gtrsim 0.37 (1.44 - l_\text{dd}/l_\perp)$. This means that the regime, where $E/N \propto (n \, l_\perp)^2$, becomes smaller, while the regime, where $E/N \propto n \, l_\perp$, becomes larger, when $l_\text{dd}/l_\perp$ approaches $l_\text{crit}/l_\perp$ from below.]

\section{Mean field theory}

\begin{figure}
\begin{center}
\includegraphics[width = \columnwidth]{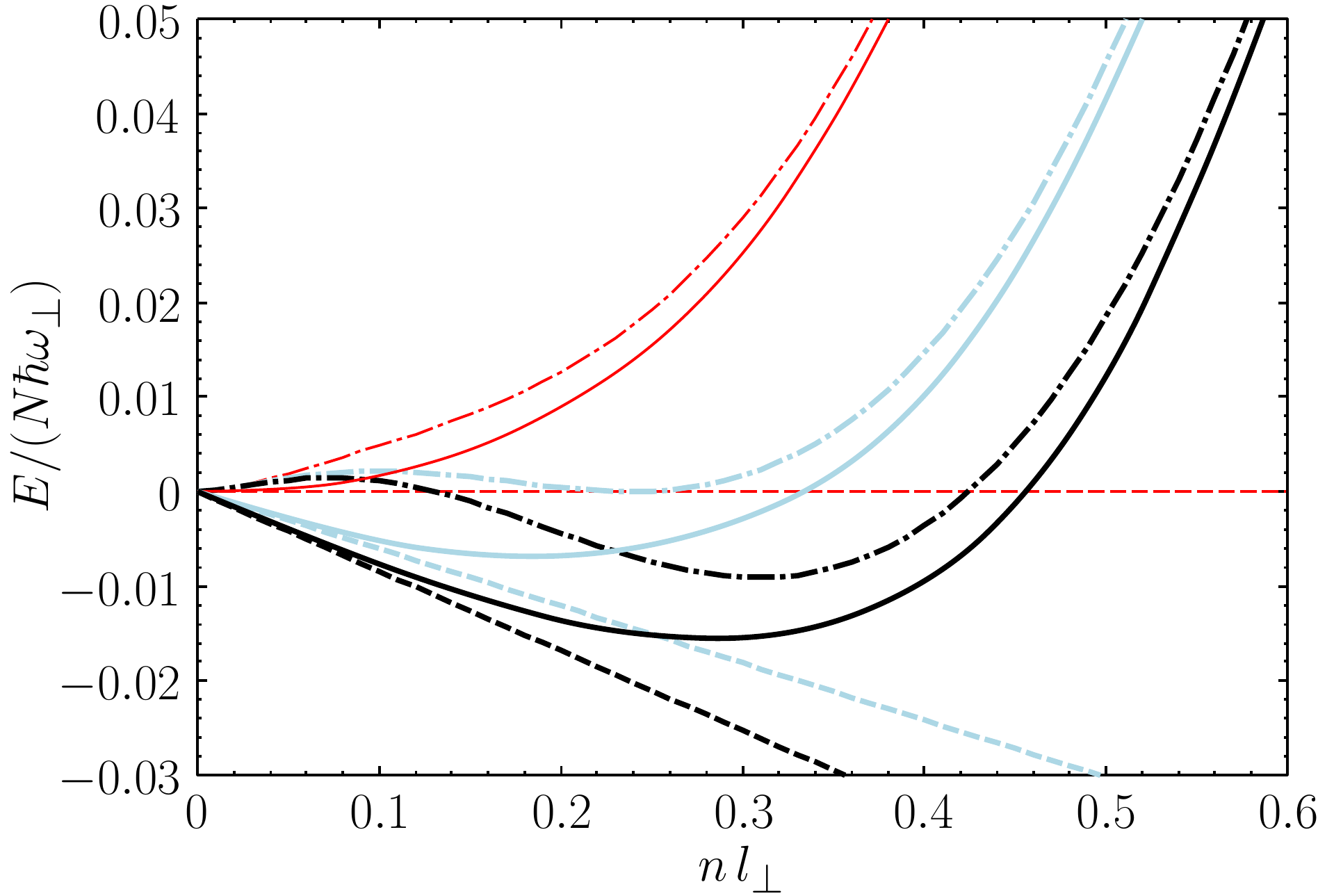}
\caption{(Color online) Many-body ground-state energy per particle $E/N$ of quasi-1D fermionic dipoles versus density $n$ for three strengths of the attractive interaction, $l_\text{dd} / l_\perp = 1.44$ [thin (red) lines], $1.77$ [thick (light-blue) lines] and $1.9$ [thick (black) lines]. Dashed lines depict the low-density limiting behavior obtained from a mapping to bosons with a contact interaction. Dash-dotted lines represent the Hartree--Fock approximation, which is reliable at high densities. Solid curves interpolate between the low- and the high-density results.}
\label{fig-Nbody}
\end{center}
\end{figure}

When the particle spacing becomes comparable to or less than the range of the potential, the mean-field approximation, which in this case is the Hartree--Fock approximation~\cite{Goral01,Miyakawa08}, becomes increasingly good. In this approximation, the energy per particle of a uniform system is given by
\begin{equation} \label{E/N}
\frac{E}{N} = \frac{\hbar^2 k_F^2}{6m} + \frac{\pi}{2 k_F} \int_{-k_F}^{k_F} \frac{dk}{2 \pi} \int_{-k_F}^{k_F} \frac{dk'}{2 \pi} \left[ \widetilde V(0) - \widetilde V(k-k') \right] \! ,
\end{equation}
where $k_F$, the Fermi wave number, is related to the density $n$ by the relation $k_F = \pi n$ and $\widetilde V(q)$ is the Fourier transform of the dipolar interaction,~(\ref{Vdd}). The first term in Eq.~(\ref{E/N}) is the kinetic energy, the term involving $\widetilde V(0)$ is the Hartree energy, and the term involving $\widetilde V(k-k')$ is the exchange energy. At high densities, $n \, l_\perp \gg l_\text{dd} / l_\perp$ the kinetic energy dominates, followed by the Hartree term, which is negative. The exchange term is positive and smaller in magnitude than the Hartree term because it involves nonzero momentum transfers $(\sim k_F)$. For small momentum transfers $q$ $\widetilde V(q) \approx \widetilde V(0) + {\cal O}(q^2 \ln q)$ and therefore at low densities the Hartree and Fock terms cancel to lowest order. The leading contribution to the interaction energy per particle is of order $n^3 \ln n$ and therefore is small compared with the kinetic energy, which varies as $n^2$. This cancellation of the Hartree and Fock terms reflects the short-range nature of the dipolar interaction in one dimension. At intermediate densities $n \, l_\perp \sim 1$ the interaction energy can be comparable to or larger in magnitude than the kinetic energy if the dipolar interaction is sufficiently strong $(l_\text{dd} \gtrsim l_\perp )$.

An explicit calculation of the interaction energy in Eq.~(\ref{E/N}) yields the energy per particle,
\begin{equation}
\frac{E}{N \hbar \omega_\perp} = \frac{\kappa_F^2}{6} - \frac{l_\text{dd}}{\pi l_\perp} \left\{ \kappa_F - \left[ I_\mathrm{arc}^{(2)} (2 \kappa_F) - \frac{I_{\ln} (\kappa_F)}{4 \kappa_F} \right] \right\} ,
\end{equation}
with $\kappa_F = k_F l_\perp$,
\begin{equation}
I_\mathrm{arc}^{(\alpha)} (x) = \int_0^\infty dw \, \arctan \left( \frac{x}{w} \right) w^\alpha e^{-w^2 / 2} ,
\end{equation}
and
\begin{equation}
I_{\ln} (x) = \int_0^\infty dw \, \ln \left( 1 + \frac{4 x^2}{w^2} \right) w^3 e^{-w^2 / 2} .
\end{equation}
In Fig.~\ref{fig-Nbody} we plot the Hartree--Fock energy for a number of values of the dipolar coupling~(dash-dotted lines). This provides an upper bound on the energy which is accurate for high densities. We also sketch as solid lines in Fig.~\ref{fig-Nbody} the expected behavior of the energy per particle, which interpolates between the linear scaling for low density, obtained from the mapping to exact results for the Bose gas, and the Hartree-Fock result for high density.

\begin{figure}
\begin{center}
\includegraphics[width = \columnwidth]{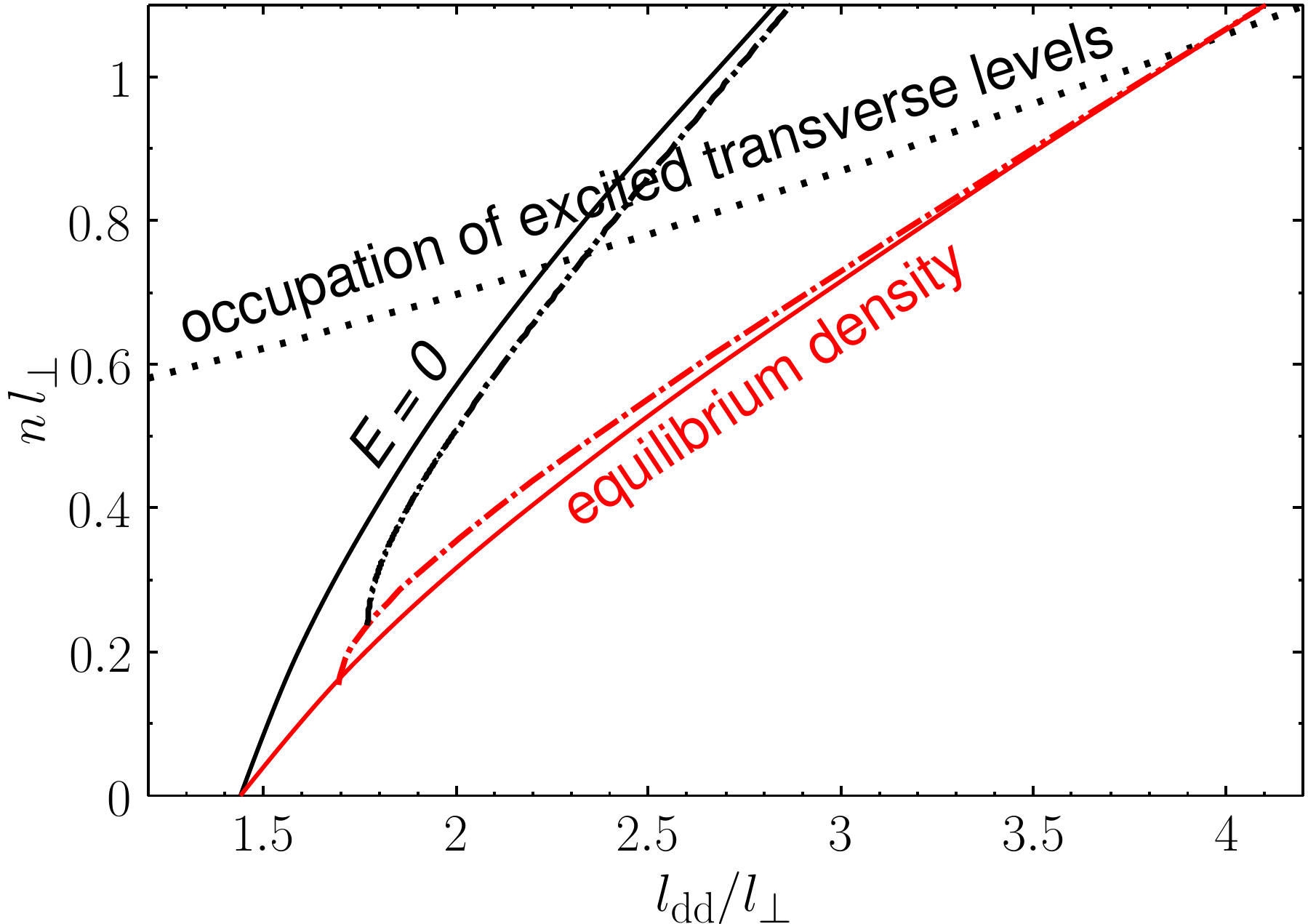}
\caption{(Color online) Phase diagram of quasi-1D fermions with attractive dipolar interactions. The solid (black) $E=0$ line interpolates between the critical point $(l_\text{dd}, n) = (1.44 \, l_\perp, 0)$ and the dash-dotted (black) Hartree--Fock $E=0$ line (see text). Above and to the left of the $E=0$ line, matter is unbound and can expand indefinitely if not confined. Below the line, matter is bound and unconfined matter will oscillate about the equilibrium density. The solid (red) curve is a sketch of the equilibrium density, which again interpolates between the critical point $(l_\text{dd}, n) = (1.44 \, l_\perp, 0)$ and the Hartree--Fock result [dash-dotted (red) line]. The onset of occupation of the first excited transverse level is shown by the dotted (black) line.}
\label{fig-phases}
\end{center}
\end{figure}

\section{Phase diagram}

On the basis of the considerations above it is possible to work out the form of the phase diagram. In Fig.~\ref{fig-phases}, we sketch the equilibrium density [solid (red) line] as a function of the dimensionless coupling strength $l_\text{dd} / l_\perp$ and the contour for zero total energy [solid (black) line]. From the exact solutions, we have shown that for $l_\text{dd} > l_\text{crit} \approx 1.44 \, l_\perp$ a low-density many-body state is bound, and therefore the contour for zero total energy must meet the coupling-constant axis for $l_\text{dd} = l_\text{crit}$. How the zero-energy contour approaches the axis depends on terms in the energy as a function of density that are of higher order than the leading one given by $g_\text{1D}^B n/2$. If they vary as $n^{1 + \gamma}$, i.e., the energy per particle is given by $g_\text{1D}^B n/2 + \alpha n^{1+\gamma}$ with $\alpha > 0$, the equilibrium density will vary as $(l_\text{dd} - l_\text{crit})^{1 / \gamma}$, since $g_\text{1D}^B \propto (l_\text{dd} - l_\text{crit})$ for small $(l_\text{dd} - l_\text{crit})$. The pressure of the system is positive above the equilibrium line. Above the $E=0$ line the system will expand to zero density if it is released, whereas it will oscillate around the equilibrium density if it is released from a density below the $E=0$ line. We also plot as dash-dotted lines the zero-energy line and the equilibrium density obtained from the Hartree--Fock approximation. The interpolations approach these lines in the high-density limit where Hartree--Fock theory is reliable. Note that since the Hartree--Fock approximation provides an upper bound on the energy, the exact zero-energy contour must lie to the left of the Hartree--Fock one.

\section{Boundary of the quasi-1D regime}

In the calculations above we have assumed that excited levels of the transverse motion play no role. In the absence of interactions, the first excited level begins to be populated when the Fermi energy becomes equal to the excitation energy of the first excited level, i.e., $\hbar^2 k_F^2 / (2m) = \hbar \omega_\perp$ or $n \, l_\perp = \sqrt{2} / \pi \approx 0.45$. When interactions are included, these criteria will change and the excited level will begin to be populated when the energy to add a particle with zero momentum in the $z$ direction in an excited level is equal to the chemical potential. In Hartree--Fock theory, this condition is
\begin{eqnarray} \label{1dboundary}
& & \mspace{-40mu} \frac{\hbar^2 k_F^2}{2m} + \int_0^{2 k_F} \frac{dk}{2 \pi} \left[ \widetilde V(0) - \widetilde V(k) \right] \nonumber \\
& & \mspace{-40mu} = \hbar \omega_\perp + 2 \int_0^{k_F} \frac{dk}{2 \pi} \left[ \widetilde V^{(0,+,0,+)}(0) - \widetilde V^{(0,+,+,0)}(k) \right] \! ,
\end{eqnarray}
where $\widetilde V^{(0,+,0,+)}$ is the Fourier transform of the direct interaction and $\widetilde V^{(0,+,+,0)}$ is the Fourier transform of the exchange interaction between two particles, one in the ground state of the transverse motion and the other in one of the two lowest excited states (see the Appendix). At low density, the sum of the Hartree and Fock contributions to the chemical potential of particles in the two bands cancel. With increasing density, the most important effect is the lack of cancellation of the Hartree and Fock terms for the ground state band. This is due to the fact that $\widetilde V(0) = 2 \widetilde V^{(0,+,0,+)}(0)$, that $\widetilde V^{(0,+,+,0)}(q)$ depends less rapidly on $q$ than does $\widetilde V(q)$, and that the integral for the excited state explores momentum transfers only up to $k_F$ while that for the ground state explores momentum transfers up to $2k_F$.

An explicit calculation of condition~(\ref{1dboundary}) yields
\begin{equation}
1 - \frac{\kappa_F^2}{2} + \frac{l_\text{dd}}{\pi l_\perp} \left[ \kappa_F - I_\mathrm{arc}^{(2)} (2 \kappa_F) + \frac{1}{2} I_\mathrm{arc}^{(4)} (\kappa_F) \right] = 0 .
\end{equation}
The dotted (black) line in Fig.~\ref{fig-phases} shows the density at which the first excited state begins to be occupied in Hartree--Fock theory. This lies above the value for noninteracting particles mainly because the Hartree term reduces the chemical potential of the lowest band. It approaches the noninteracting result $n \, l_\perp \approx 0.45$ for $l_\text{dd} / l_\perp \rightarrow 0$.

\section{Experimental considerations}

In future experiments with polar molecules it should be possible to create self-bound fermionic clusters. Consider, for example, the case of $^{23}$Na$^{40}$K molecules, which are particularly interesting since they are chemically nonreactive~\cite{Zuchowski10} and possess a dipole moment of $2.73$ Debye (D)~\cite{Aymar05} if fully polarized. Partially polarized molecules acquire an induced electric dipole moment of 1 D in a modest external electric field ($\sim$10\,kV\,/\,cm), which corresponds to a dipole length $l_\text{dd} \approx 0.94 \mu$m. The regime of self-bound states, $l_\text{dd} \gtrsim 1.44 \, l_\perp$, then corresponds to transverse trapping frequencies $\omega_\perp \gtrsim 2 \pi \times 377 \,$Hz. Self-bound clusters could be detected by absorption imaging, since, in the presence of a trapping potential in the $z$-direction, the size of a cloud of atoms would decrease very rapidly as the coupling strength passes through the critical value. Also, when the cloud of atoms is bound, the cloud will not expand indefinitely when the trapping in the $z$-direction is turned off.

\section{Concluding remarks}

In this paper we have explored the phase diagram of a quasi-1D dipolar Fermi gas with attractive interactions. This phase diagram, characterized by the appearance of a self-bound many-body state, has been obtained by a combination of techniques: exact mappings for 1D systems with contact interactions, which enables one to pin down properties at low densities, and mean-field theory, which is reliable at high densities. Our work shows that dipolar fermionic gases make possible an experimental realization of generalized contact interactions under conditions that can be achieved in current experiments. Our calculations also bring out the important role played by the transverse extent of the system. A system with nonzero $l_\perp$ behaves very differently from a purely 1D system~\cite{Girardeau12}, since the singularity in the purely 1D dipolar interaction is absent and thereby collapse is hindered. A topic for future work is how virtual excitation of transverse excited states influences the effective 1D coupling constant. As long as no real excitations are present in higher transverse levels, the main effect will be a renormalization of the interaction scale, since the effective 1D coupling constant will not be equal to $l_\text{dd}$.

\section*{\uppercase{Acknowledgments}}

We thank J.~C.~Cremon and T.~Vekua for helpful discussions. This work was supported in part by the DFG~(SA1031/6), the Cluster of Excellence QUEST, the German-Israeli Foundation, the Carlsberg Foundation, the ESF POLATOM network, the Swedish Research Council, and the Nanometer Structure Consortium at Lund University.

\section*{\uppercase{Appendix: Matrix elements of the interaction for particles in excited transverse levels}}

The effective 1D matrix elements of the dipole potential between an initial state with two particles in transverse states $\phi_\gamma (x,y)$ and $\phi_\delta (x,y)$ and a final state with particles in states $\phi_\alpha (x,y)$ and $\phi_\beta (x,y)$ are given by
\begin{eqnarray}
& & \mspace{-50mu} V^{(\alpha,\beta,\gamma,\delta)} (z - z') \nonumber \\
& & \mspace{-50mu} = \int d^2 \rho \, d^2 \rho' \, \phi_\alpha^* (\vec \rho \,) \phi_\gamma (\vec \rho \,) \phi_\beta^* (\vec \rho\,') \phi_\delta (\vec \rho\,') V (\vec r - \vec r\,') .
\end{eqnarray}
Here, we take the states of the transverse motion to be angular momentum and energy eigenstates of an axially symmetric harmonic oscillator. The ground state and the two lowest excited states are given by
\begin{eqnarray}
\phi_0 (x,y) & = & \frac{1}{\sqrt{\pi} l_\perp} e^{-(x^2 + y^2) / (2 l_\perp^2)} , \\
\phi_\pm (x,y) & = & \frac{1}{\sqrt{\pi} l_\perp^2} (x \pm {\rm i} y) e^{-(x^2 + y^2) / (2 l_\perp^2)} ,
\end{eqnarray}
and the 3D interaction potential for dipoles oriented along the $z$ axis is
\begin{equation}
V (\vec r\,) = \frac{d^2}{r^3} \left( 1 - \frac{3 z^2}{r^2} \right).
\end{equation}
The matrix elements are best calculated in Fourier~($k$) space, and the Fourier transform of the 3D dipolar interaction potential is given by
\begin{equation}
\widetilde V (\vec k) = \frac{4 \pi d^2}{3} \left( \frac{3 k_z^2}{k^2} - 1 \right) .
\end{equation}
The effective 1D matrix elements acquire the form
\begin{eqnarray}
& & V^{(\alpha,\beta,\gamma,\delta)} (z) = \left[ \frac{2 d^2}{3 \pi} \int d^2 k_\rho \, \widetilde{\phi_\alpha^* \phi_\gamma} (-\vec k_\rho) \widetilde{\phi_\beta^* \phi_\delta} (\vec k_\rho) \right] \delta (z) \nonumber \\
& & - \frac{d^2}{2 \pi} \int d^2 k_\rho \, k_\rho \widetilde{\phi_\alpha^* \phi_\gamma} (-\vec k_\rho) \widetilde{\phi_\beta^* \phi_\delta} (\vec k_\rho) e^{-k_\rho |z|} ,
\end{eqnarray}
where $\widetilde{\phi_\alpha^* \phi_\beta}$ is the  Fourier transform of $\phi_\alpha^*\phi_\beta$ with respect to the transverse coordinates. In the text, we use the matrix elements
\begin{equation}
V(z) \equiv V^{(0,0,0,0)} (z) = -\frac{d^2}{l_\perp^3} \int_0^\infty d w \, w^2 e^{- w^2 / 2 - w |z| / l_\perp} ,
\end{equation}
\begin{equation}
V^{(0,+,0,+)} (z) = -\frac{d^2}{l_\perp^3} \int_0^\infty d w \, w^2 \biggl( 1 - \frac{w^2}{4} \biggr) e^{- w^2 / 2 - w |z| / l_\perp} ,
\end{equation}
and
\begin{equation}
V^{(0,+,+,0)} (z) = -\frac{d^2}{4 l_\perp^3} \int_0^\infty d w \, w^4 e^{- w^2 / 2 - w |x| / l_\perp} ,
\end{equation}
where we have omitted contact terms, since they play no role in our calculations due to the cancellation of the direct and exchange contributions. (In contrast, the contact term plays an important role in dipolar Bose gases \cite{Bartolo13}.)

\bibliographystyle{prsty}

\begin{thebibliography}{}

\bibitem{Cheon99} T. Cheon and T. Shigehara, \doi{10.1103/PhysRevLett.82.2536}{Phys. Rev. Lett. {\bf 82}, 2536 (1999)}.

\bibitem{Granger04} B. E. Granger and D. Blume, \doi{10.1103/PhysRevLett.92.133202}{Phys. Rev. Lett. {\bf 92}, 133202 (2004)}.

\bibitem{Lieb63} E. H. Lieb and W. Liniger, \doi{10.1103/PhysRev.130.1605}{Phys. Rev. {\bf 130}, 1605 (1963)}.

\bibitem{McGuire64} J. B. McGuire, \doi{10.1063/1.1704156}{J. Math. Phys. {\bf 5}, 622 (1964)}.

\bibitem{Giorgini08} S. Giorgini, L. P. Pitaevskii, and S. Stringari, \doi{10.1103/RevModPhys.80.1215}{Rev. Mod. Phys. {\bf 80}, 1215 (2008)}.

\bibitem{Baranov08} M. A. Baranov, \doi{10.1016/j.physrep.2008.04.007}{Phys. Rep. {\bf 464}, 71 (2008)}.

\bibitem{Lahaye09} T. Lahaye, C. Menotti, L. Santos, M. Lewenstein, and T. Pfau, \doi{10.1088/0034-4885/72/12/126401}{Rep. Prog. Phys. {\bf 72}, 126401 (2009)}.

\bibitem{Baranov12} M. A. Baranov, M. Dalmonte, G. Pupillo, and P. Zoller, \doi{10.1021/cr2003568}{Chem. Rev. {\bf 112}, 5012 (2012)}.

\bibitem{Ni08} K.-K. Ni, S. Ospelkaus, M. H. G. de Miranda, A. Pe'er, B.~Neyenhuis, J. J. Zirbel, S. Kotochigova, P. S. Julienne, D.~S.~Jin, and J. Ye, \doi{10.1126/science.1163861}{Science {\bf 322}, 231 (2008)}.

\bibitem{Chotia12} A. Chotia, B. Neyenhuis, S. A. Moses, B. Yan, J. P. Covey, M.~Foss-Feig, A. M. Rey, D. S. Jin, and J. Ye, \doi{10.1103/PhysRevLett.108.080405}{Phys. Rev. Lett. {\bf 108}, 080405 (2012)}.

\bibitem{Wu12} C.-H. Wu, J. W. Park, P. Ahmadi, S. Will, and M. W. Zwierlein, \doi{10.1103/PhysRevLett.109.085301}{Phys. Rev. Lett. {\bf 109}, 085301 (2012)}.

\bibitem{Heo12} M.-S. Heo, T. T. Wang, C. A. Christensen, T. M. Rvachov, D. A. Cotta, J.-H. Choi, Y.-R. Lee, and W. Ketterle, \doi{10.1103/PhysRevA.86.021602}{Phys. Rev. A {\bf 86}, 021602(R) (2012)}.

\bibitem{Sinha07} S. Sinha and L. Santos, \doi{10.1103/PhysRevLett.99.140406}{Phys. Rev. Lett. {\bf 99}, 140406 (2007)}.

\bibitem{Deuretzbacher10} F. Deuretzbacher, J. C. Cremon, and S. M. Reimann, \doi{10.1103/PhysRevA.81.063616}{Phys. Rev. A {\bf 81}, 063616 (2010)}; \doi{10.1103/PhysRevA.87.039903}{{\bf 87}, 039903(E) (2013)}.

\bibitem{Bartolo13} N. Bartolo, D. J. Papoular, L. Barbiero, C. Menotti, and A. Recati, \doi{10.1103/PhysRevA.88.023603}{Phys. Rev. A {\bf 88}, 023603 (2013)}.

\bibitem{Calogero75} F. Calogero and A. Degasperis, \doi{10.1103/PhysRevA.11.265}{Phys. Rev. A {\bf 11}, 265 (1975)}.

\bibitem{Goral01} K. G{\'o}ral, B.-G. Englert, and K. Rz\c{a}\.{z}ewski, \doi{10.1103/PhysRevA.63.033606}{Phys. Rev. A {\bf 63}, 033606 (2001)}.

\bibitem{Miyakawa08} T. Miyakawa, T. Sogo, and H. Pu, \doi{10.1103/PhysRevA.77.061603}{Phys. Rev. A {\bf 77}, 061603(R) (2008)}.

\bibitem{Zuchowski10} P. S. Zuchowski and J. M. Hutson, \doi{10.1103/PhysRevA.81.060703}{Phys. Rev. A {\bf 81}, 060703(R) (2010)}.

\bibitem{Aymar05} M. Aymar and O. Dulieu, \doi{10.1063/1.1903944}{J. Chem. Phys. {\bf 122}, 204302 (2005)}.

\bibitem{Girardeau12} M. D. Girardeau and G. E. Astrakharchik, \doi{10.1103/PhysRevLett.109.235305}{Phys. Rev. Lett. {\bf 109}, 235305 (2012)}.

\end{thebibliography}

\end{document}